\documentclass[aps,prb,twocolumn,superscriptaddress,longbibliography]{revtex4-2}
\usepackage{epsfig,amsopn}
\usepackage{graphicx}
\usepackage{physics}
\usepackage{color}
\usepackage{amsmath,amssymb}
\usepackage{enumerate}
\newcommand\bea{\begin{eqnarray}}
\newcommand\eea{\end{eqnarray}}
\newcommand\beq{\begin{equation}}
\newcommand\eeq{\end{equation}}

\def\nn{\nonumber}
\def\f{\frac}
\def\al{\alpha}

\def\si{\sigma}
\def\Do{\partial}

\def\la{\langle}
\def\ra{\rangle}
\def\ua{\uparrow}
\def\da{\downarrow}

\def\th{\theta}

\begin{document}
\title{Nonreciprocal transverse currents in Rashba metal junctions under out-of-plane Zeeman fields}
\author{Megha Bera}
\affiliation{School of Physics, University of Hyderabad, Prof. C. R. Rao Road, Gachibowli, Hyderabad-500046, India}
\author{Bijay Kumar Sahoo}
\affiliation{School of Physics, University of Hyderabad, Prof. C. R. Rao Road, Gachibowli, Hyderabad-500046, India}
\author{ Abhiram Soori}
\email{abhirams@uohyd.ac.in}
\affiliation{School of Physics, University of Hyderabad, Prof. C. R. Rao Road, Gachibowli, Hyderabad-500046, India}

\begin{abstract}
We study charge transport across a junction between a normal metal and a Rashba metal in the presence of a Zeeman field applied to the spin--orbit coupled region. While an out-of-plane Zeeman field does not generate a transverse response in a homogeneous Rashba system, we show that such a junction exhibits a finite transverse conductivity that is inherently nonreciprocal, i.e., it depends on the direction of the applied bias. We demonstrate that this effect originates from the breaking of the $k_y \to -k_y$ symmetry of the Hamiltonian in the presence of the Zeeman field, which prevents cancellation of transverse current contributions from opposite transverse momenta. We further show that evanescent modes in the spin--orbit coupled region play a crucial role by carrying a finite spin polarization that gives rise to a transverse current localized near the junction. The transverse conductivity exhibits a peak at an energy scale set by the Zeeman field, displays distinct behavior for opposite bias directions, and shows spatial dependence governed by the nature of the contributing modes. We also identify bound states at the junction for attractive barrier strengths, which enhance conductivity when their energies lie near the transport window. Our results reveal a mechanism for nonreciprocal transverse charge transport in Rashba systems without requiring in-plane magnetic fields or ferromagnetic contacts, and should be experimentally accessible in semiconductor heterostructures.
\end{abstract}

\maketitle

\section{Introduction}

Electron transport in two-dimensional electron gases (2DEGs) with spin--orbit coupling (SOC) has been extensively studied in condensed matter physics. In such systems, structural inversion asymmetry gives rise to Rashba SOC~\cite{rashba60,bychkov84}, leading to spin-momentum locking and a variety of spin-dependent transport phenomena~\cite{pepper99}. Materials hosting Rashba SOC are commonly referred to as Rashba metals.

In this work, we demonstrate that a junction between a normal metal (NM) and a Rashba metal exhibits a finite and \emph{nonreciprocal} transverse conductivity under an out-of-plane Zeeman field, despite the absence of any Fermi surface shift or intrinsic anisotropy. The transverse response depends on the direction of the applied bias and shows qualitatively different behavior for transport from left to right and from right to left.

Transverse charge responses in Rashba systems are typically associated with mechanisms that break symmetry in momentum space. For example, in junctions involving a normal metal and a Rashba metal, an in-plane Zeeman field shifts the Fermi surface in the transverse direction and generates a finite transverse conductivity~\cite{soori2021}, an effect known as the planar Hall effect~\cite{tang03,Li10,Roy10}. Similarly, in Datta--Das spin transistor setups~\cite{dattadas,sarkar2019,sarkar2020}, where a Rashba metal is connected to ferromagnetic leads, transverse responses can arise when the spin polarization is oriented along specific directions~\cite{Sahoo2023}. More generally, transverse transport in nonmagnetic metals can result from anisotropy of the Fermi surface~\cite{soori2026trans} or from geometric anisotropy of the system~\cite{song98,sarkar2025}. In contrast, an out-of-plane Zeeman field does not shift the Fermi surface, and therefore a transverse response is not expected in homogeneous Rashba systems.

We show that the transverse response in the present system arises from a distinct mechanism that relies on interfacial scattering. In particular, evanescent modes in the spin--orbit coupled region carry a finite spin polarization and contribute asymmetrically to transverse transport. In the presence of the Zeeman field, the Hamiltonian no longer possesses $k_y \to -k_y$ symmetry, and as a result, the transverse current contributions from opposite transverse momenta do not cancel. The interplay of this symmetry breaking with the junction geometry leads to a finite and inherently nonreciprocal transverse conductivity.

\begin{figure}[htb]
\includegraphics[width=8cm]{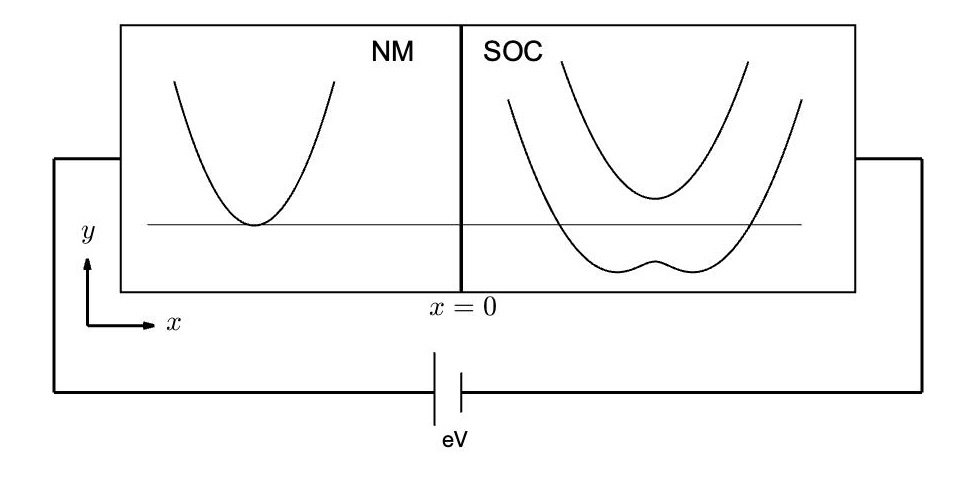}
\caption{Schematic of the system: a junction between a normal metal (NM) and a spin--orbit coupled (SOC) region subjected to an out-of-plane Zeeman field. The dispersion relations are drawn in NM and SOC regions. }
\label{fig:schem}
\end{figure}

We analyze this effect in detail by computing the longitudinal and transverse conductivities as functions of bias and system parameters. We show that the transverse conductivity exhibits a peak at an energy scale set by the Zeeman field, displays qualitatively different behavior for opposite bias directions, and shows distinct spatial dependence depending on whether the contributing modes are propagating or evanescent. We also demonstrate that bound states can form at the junction for attractive barrier strengths, leading to enhanced conductance when their energies lie close to the transport window. A schematic of the setup is shown in Fig.~\ref{fig:schem}.

\section{Details of calculation}

The system is described by the Hamiltonian  
\beq 
H=
\begin{cases}
-\f{\hbar^2}{2m_n}\big(\f{\Do^2}{\Do x^2}+\f{\Do^2}{\Do y^2}\big)\si_0~~\text{for}~~x<0,  \\ 
-\f{\hbar^2}{2m_s}\big(\f{\Do^2}{\Do x^2}+\f{\Do^2}{\Do y^2}\big)\si_0\\  ~~~~~~+i\al\big( \si_x\f{\Do}{\Do y} -\si_y\f{\Do}{\Do x} \big) +b\si_z ~~~\text{for}~~x>0,
\end{cases}
\eeq
where $m_n$ ($m_s$) is the effective mass in normal metal (SOC) region, $\alpha$ is the strength of Rashba spin--orbit coupling, and $b$ is the Zeeman energy. The dispersion relation in the NM region is $E=\hbar^2(k_x^2+k_y^2)/2m_n$ for both spin species, while in the SOC region it is $E=\hbar^2(k_x^2+k_y^2)/2m_s\pm\sqrt{\al^2(k_x^2+k_y^2)+b^2}$.

We emphasize that the out-of-plane Zeeman splitting in our setup does not originate from an external macroscopic magnetic field, but rather models a proximity-induced exchange field created by an adjacent ferromagnetic insulator~\cite{Zutic2019, Wei2013}. Because this magnetic proximity effect is governed by the short-range atomic overlap of wavefunctions at the interface, the induced exchange field strictly vanishes outside the covered region, naturally yielding the step-like spatial profile adopted in our model.

For transport calculations, appropriate boundary conditions are required. Imposing current conservation at the junction yields  
\bea
\psi_{0^-} &=& c\psi_{0^+}, \nn \\ 
c\Big[\f{\Do_x\psi}{m_n}+\f{q_0}{m_n}\psi\Big]_{0^-} &=& \Big[\f{\Do_x\psi}{m_s}+i\f{\alpha\sigma_y\psi}{\hbar}\Big]_{0^+} ,~\label{eq:bc}
\eea
where $c$ is a real dimensionless parameter. 
Here, $q_0$ characterizes the strength of a delta-function potential barrier at the interface, entering the Hamiltonian as $V(x) = (\hbar^2 q_0/m) \delta(x)$. The dimensionless parameter $c$ describes the effective interfacial coupling. Specifically, in the absence of spin-orbit coupling ($\alpha=0$), $c$ reduces to the ratio $c = t'/t$, where $t'$ and $t$ are the interfacial and bulk hopping amplitudes in an equivalent tight-binding model~\cite{soori23scat}. For the remainder of this work, we set $c=1$, which recovers the standard wavefunction continuity condition across the junction.

We calculate the longitudinal and transverse responses of the system to a bias applied from left to right and from right to left separately, as described in the following subsections.

\subsection{Scattering of electrons incident from left to right}

\begin{figure}[htb]
\includegraphics[width=7cm]{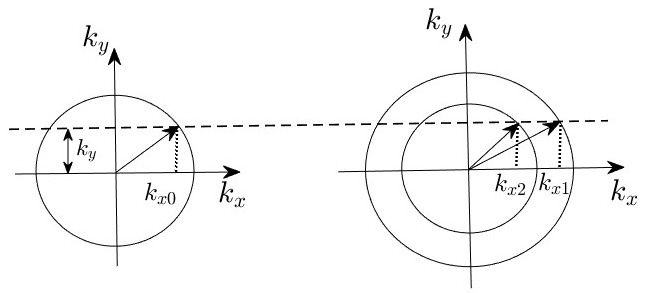}
\caption{Different wavenumbers that enter the calculation are shown. Circles on the left (right) show the constant energy contours of normal metal (SOC region). }\label{fig:k}
\end{figure}

For a $\si$-spin electron incident from $x<0$ onto the junction with energy $E$ and angle of incidence $\th$, the scattering eigenfunction has the form $\psi(x)e^{ik_yy}$, where  
\beq 
\psi_{\si}(x) =
\begin{cases}
e^{ik_{x0}x}\ket{\si}+r_{\ua\si}e^{-ik_{x0}x}\ket{\ua}+r_{\da\si}e^{-ik_{x0}x}\ket{\da}, \\~~~~~~~~~~~~~~~~~~~~~~~~~~~~~~~~~\text{for}~~x<0\\
t_{1\si}e^{ik_{x1}x}\ket{1}+t_{2\si}e^{ik_{x2}x}\ket{2},~~\text{for}~~x>0,
\end{cases}\label{eq:psi}
\eeq
with $\si=\ua,\da$, $\ket{\ua}=[1,0]^T$, and $\ket{\da}=[0,1]^T$. The wavenumbers are given by $k_{x0}=\cos{\th}\sqrt{2m_nE}/\hbar$ and $k_y=\sin{\th}\sqrt{2m_nE}/\hbar$. The quantities $k_{x1}$ and $k_{x2}$ are the two solutions for $k_x$ obtained from the dispersion relation in the SOC region, chosen such that each corresponds either to a right-moving mode or to an evanescent (decaying) mode. The transverse momentum $k_y$ is conserved across the junction due to translational invariance along the $y$-direction. The spinors $\ket{1}$ and $\ket{2}$ are the eigenspinors of the SOC-region Hamiltonian corresponding to $k_x=k_{x1}$ and $k_x=k_{x2}$, respectively. The wavenumbers $k_{x0}$, $k_{x1}$, $k_{x2}$ and $k_{y}$ are pictorially shown in Fig.~\ref{fig:k}. 

Applying the boundary conditions in Eq.~\eqref{eq:bc}, the scattering coefficients in Eq.~\eqref{eq:psi} can be determined. These coefficients are then used to compute the differential conductivities $G_{xx}=dI_x/dV$ and $G_{yx}=dI_y/dV$, where $I_x$ ($I_y$) is the current density along the $x$ ($y$) direction and $V$ is the applied bias. The conductivities are given by  
\bea
G_{xx} &=& \f{em_n}{4\pi^2\hbar^2}\int_{-\pi/2}^{\pi/2}d\th\, I_{x}, \nn \\ 
G_{yx} &=& \f{em_n}{4\pi^2\hbar^2}\int_{-\pi/2}^{\pi/2}d\th\, I_{y}, \label{eq:condu}
\eea
where  
\bea
I_{x} &=& \f{e\sqrt{2E}}{\sqrt{m_n}}\cos{\th}\left(2-\sum_{\si,\si'}|r_{\si\si'}|^2\right), \nn \\
I_{y} &=&
\begin{cases}
{e\hbar k_y}\sum_{\si=\ua,\da}\psi_{\si}^{\dagger}(x)\psi_{\si}(x)/m_n,~~~{\rm for ~~}x<0,\\~~~\\
e\sum_{\si=\ua,\da}\Big[{\hbar k_y}\psi_{\si}^{\dagger}(x)\psi_{\si}(x)/m_s \nn \\~~~~~~~~ -{\al}\psi_{\si}^{\dag}(x)\si_x\psi_{\si}(x)/\hbar\Big], 
{\rm ~~~~for~~ }x>0 .
\end{cases} \label{eq:current}
\eea

\subsection{Scattering of electrons incident from right to left}

An electron can also be incident from the SOC region onto the junction. For $E>b$, the electron may originate from either of the two bands, whereas for $0<E<b$, only one band supports propagating states (the other corresponding to a complex wavevector).

The scattering eigenfunction for an electron incident in band $j$ with energy $E$ and angle $\th$ has the form $\psi_j(x)e^{ik_yy}$, where  
\beq 
\psi_j(x) =
\begin{cases}
e^{-ik_{xj}x}\ket{j_-} + r'_{1j}e^{ik_{x1}x}\ket{1}+r'_{2j}e^{ik_{x2}x}\ket{2}, \\ 
\text{for}~~x>0,\\
t'_{\ua j}e^{-ik_{xn}x}\ket{\ua}+t'_{\da j}e^{-ik_{xn}x}\ket{\da}, \\ 
\text{for}~~x<0. 
\end{cases}\label{eq:psirl}
\eeq

Here, $k_y=k_j\sin{\th}$, where $k_j$ is a solution of the SOC dispersion $E=\hbar^2k^2/2m_s+s_j\sqrt{\al^2k^2+b^2}$ with $s_j=(-1)^j$. For real $k_j$, we define $k_{xj}=k_j\cos{\th}$. The spinor $\ket{j_-}$ denotes the normalized eigenstate corresponding to a left-moving mode in band $j$. The wavenumbers $k_{x1},\,k_{x2}$ correspond to right-moving or evanescent modes in bands $1,\,2$ with the same energy $E$ and transverse momentum $k_y$. The wavenumber $k_{xn}$ corresponds to the NM region with the same $k_y$ as that of the incident wave and is chosen to be positive when real, or to have a positive imaginary part when complex.
The roots $k_1$ and $k_2$ are chosen such that $k_1$ is always real (for $E>0$) and positive. When $k_2$ is also real (which happens for $E>b$), $k_1>k_2>0$. Note that $\psi_j(x)$ is defined only when $k_j$ is real.

The scattering coefficients $r'_{j'j}$ and $t'_{\si j}$ are obtained using Eq.~\eqref{eq:bc}. The corresponding current densities are  
\bea
I_{x,j} &=& e\sqrt{\f{2E}{m_n}}\cos{\th}\left(|t'_{\ua,j}|^2+|t'_{\da,j}|^2\right), \nn \\
I_{y,j} &=&
\begin{cases}
{e\hbar k_y}\psi_{j}^{\dagger}(x)\psi_{j}(x)/m_n,~~~{\rm for ~~}x<0,\\
e\Big[{\hbar k_y}\psi_{j}^{\dagger}(x)\psi_{j}(x)/m_s -{\al}\psi_{j}^{\dag}(x)\si_x\psi_{j}(x)/\hbar\Big], \\ 
~~~~~~~~~~~~~~~~~~~~~~~~~~~~~{\rm for~~ }x>0 ,
\end{cases}
\eea 
Note that when $k_j$ is not real, both $I_{x,j}$ and $I_{y,j}$ vanish.
The conductivities are then given by  
\bea 
G_{xx} &=& \sum_{j=1,2}\f{e}{4\pi^2\big(\f{\hbar^2}{m_s}+s_j\f{\al^2}{\sqrt{\al^2k_j^2+b^2}}\big)}\int_{-\pi/2}^{\pi/2}d\th\, I_{x,j}, \nn \\ 
G_{yx} &=& \sum_{j=1,2}\f{e}{4\pi^2\big(\f{\hbar^2}{m_s}+s_j\f{\al^2}{\sqrt{\al^2k_j^2+b^2}}\big)}\int_{-\pi/2}^{\pi/2}d\th\, I_{y,j} .
\eea

 \section{Results}
 \subsection{Bias dependence and nonreciprocal transverse response}
 We first investigate the longitudinal and transverse conductivities as functions of the applied bias, focusing on the emergence of a nonreciprocal transverse response. To ensure experimental relevance, our transport calculations utilize parameters typical for an InAs two-dimensional electron gas (2DEG)~\cite{kochan2023}. To avoid spurious interfacial scattering arising from a large Fermi momentum mismatch, we model the junction within a single continuous InAs heterostructure. While InAs possesses intrinsic spin-orbit coupling, its magnitude is highly tunable via local electrostatic gating~\cite{matsuyama2000,shojaei2016,hatke2017}. Thus, we treat the un-gated (or symmetrically gated) section as an effective normal metal with vanishing spin-orbit coupling, while the asymmetrically gated region hosts the Rashba interaction. Throughout our calculations, we set the effective masses $m_s=m_n=0.1m_e$, the Rashba parameter $\alpha=2\alpha_0$, and the Zeeman splitting $b=0.25E_F$. Here, $m_e$ is the bare electron mass, the Fermi energy is $E_F=1.5$\,meV, and the  spin-orbit energy scale is $\alpha_0=\hbar\sqrt{E_F/m_e}=10.6$\,meV\,nm.
 
 \begin{figure}[htb]
\includegraphics[width=8cm]{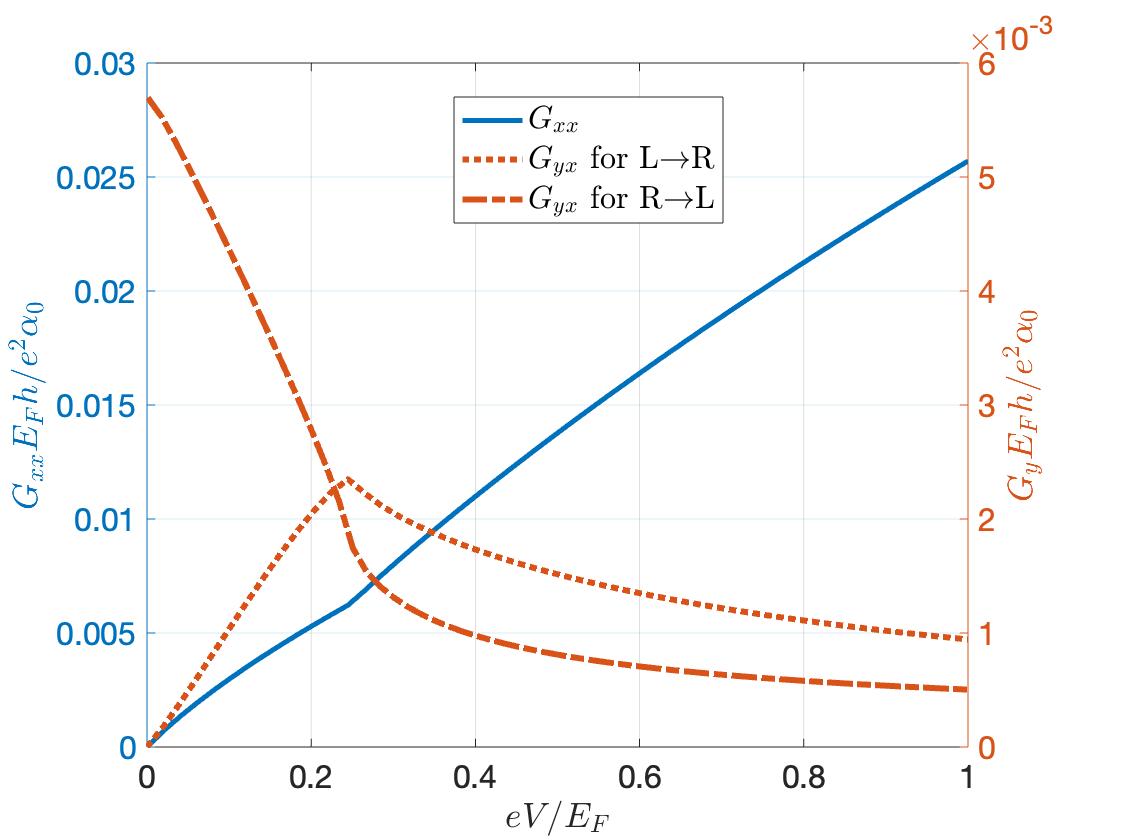}
\caption{Longitudinal ($G_{xx}$, on the left axis) and transverse ($G_{yx}$, on the right axis) conductivities as functions of bias. While $G_{xx}$ is identical for left-to-right (L$\to$R) and right-to-left (R$\to$L) bias, $G_{yx}$ differs for the two directions. Note the strong non-reciprocity in $G_{yx}$. Parameters: $c=1$, $q_0=0.5E_F/\al_0$, $\al=2\al_0$, $b=0.25E_F$, evaluated at $x=0^{+}$.}
\label{fig:GvsE}
\end{figure}

Figure~\ref{fig:GvsE} shows the longitudinal and transverse differential conductivities as functions of bias for transport from left to right, and right to left. Both conductivities vanish at zero bias and increase with increasing bias $eV$ for bias from left to right. This behavior originates from the vanishing group velocity on the NM side at zero energy. As the bias increases, the velocities of propagating modes increase in both regions. The longitudinal conductivity $G_{xx}$, which is the same for biases in the two directions, exhibits a change in slope at $eV=b$. This reflects the onset of contributions from both bands in the SOC region for $eV>b$, whereas only one band contributes for $eV<b$. 

In contrast, the transverse conductivity $G_{yx}$ exhibits a peak near $eV=b$ and decreases at higher bias. Although the dispersions on both sides are symmetric under $k_y\to -k_y$, the transverse response is nonzero. In the limit $b\to 0$, one recovers $G_{yx}=0$ due to exact cancellation between contributions from $k_y$ and $-k_y$.

\subsection{Origin of transverse current}
The SOC Hamiltonian in momentum space is 
\bea
h_{\vec k} &=& \f{\hbar^2(k_x^2+k_y^2)}{2m_s}\si_0 + \al(k_x\si_y-k_y\si_x) +b\si_z.
\eea
For $b=0$, the Hamiltonian is invariant under the transformation $k_y \to -k_y$ combined with a spin rotation. In particular, under $k_y \to -k_y$, the term $\al(-k_y\si_x)$ changes sign, but this can be compensated by a unitary transformation (by acting $\si_y$ on either sides), which flips the sign of $\si_x$ while leaving the rest of the Hamiltonian unchanged. As a result, states with $(k_x,k_y)$ and $(k_x,-k_y)$ are related by symmetry and contribute equally and oppositely to the transverse current, leading to an exact cancellation and hence $G_{yx}=0$. 

In contrast, when $b \neq 0$, the Zeeman term $b\si_z$ does not remain invariant under the same spin rotation that compensates $k_y \to -k_y$. Consequently, the Hamiltonian is no longer symmetric under $k_y \to -k_y$, and the contributions to the transverse current from $k_y$ and $-k_y$ do not cancel. This lack of cancellation results in a finite transverse conductivity.

We also compute $\la\si_x\ra$, averaged over the eigenstates, and find that it follows the same dependence on energy as $G_{yx}$. Since $\si_x$ couples directly to $k_y$ in the Hamiltonian, this provides an alternative interpretation of the transverse response.

At large bias, $E\gg b$, the Zeeman term becomes negligible and the transverse conductivity approaches zero. At zero bias, transport vanishes due to zero group velocity on the NM side. Consequently, $G_{yx}$ must exhibit a peak at an intermediate energy scale set by $b$.

For transport from right to left, $G_{yx}$ is finite even at zero bias. This arises from complete reflection of incident electrons, which nevertheless acquire a finite transverse current due to the strong influence of the Zeeman field near the gap.

\subsection{Spatial dependence of transverse conductivity}
\begin{figure}[htb]
\includegraphics[width=8cm]{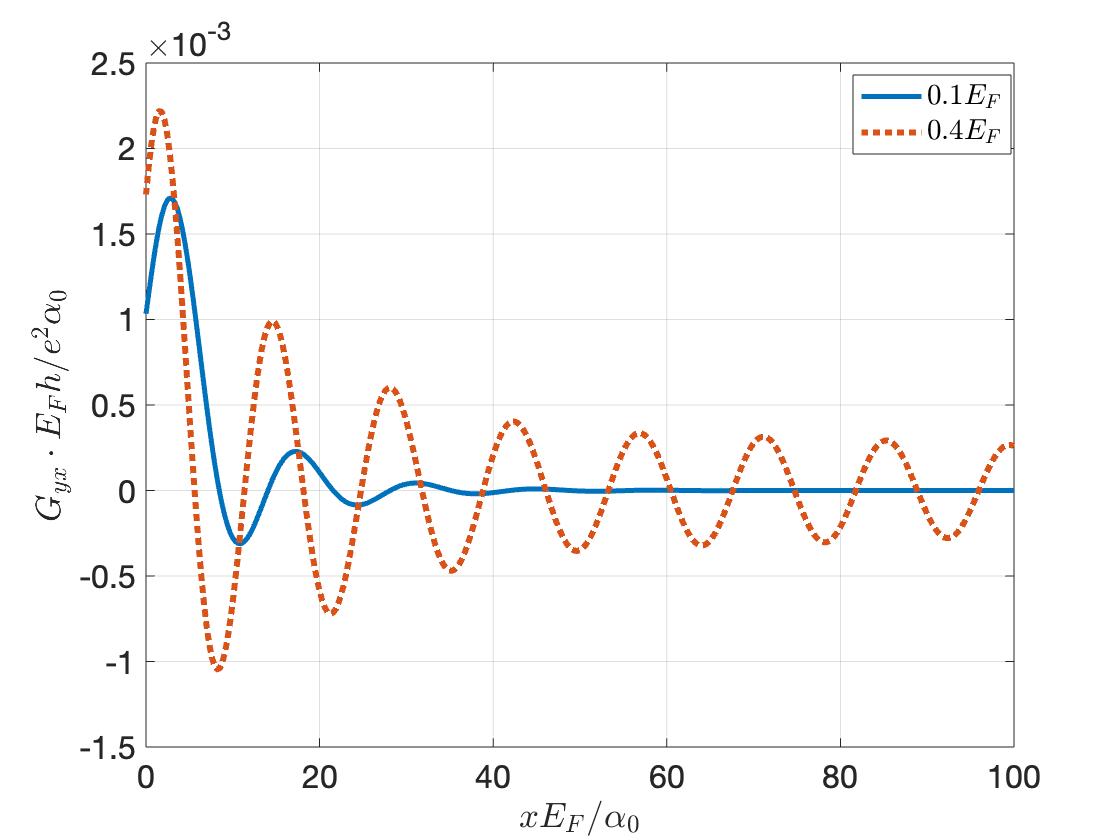}
\caption{Transverse conductivity $G_{yx}$ as a function of position $x$ for left-to-right bias at $eV=0.1E_F$ (solid line) and $eV=0.4E_F$ (dashed line). Other parameters are the same as in Fig.~\ref{fig:GvsE}.}
\label{fig:vsx}
\end{figure}

The transverse conductivity depends on the spatial location at which it is evaluated. It vanishes in the NM region. As shown in Fig.~\ref{fig:vsx}, for $eV<b$, it decays to zero deep inside the SOC region, whereas for $eV>b$, it exhibits oscillations around zero.

This behavior can be understood from the structure of the SOC eigenstates. For real $k_{xj}$ (typically for $E>b$), contributions from $(k_y,-k_y)$ cancel at the level of individual bands, but interference between different bands leads to oscillatory behavior through factors of $e^{\pm i(k_{x1}-k_{x2})x}$. 

For complex $k_{xj}$, the corresponding modes in the SOC region are evanescent, with wavefunctions that decay exponentially away from the interface. In this case, the eigenspinor has the form $\ket{j}\propto [\al(k_y+ik_{xj}),~b-E+\hbar^2(k_{xj}^2+k_y^2)/2m_s]^T$, where $k_{xj}$ is complex. The presence of the factor $i k_{xj}$ in the first component implies that both the real and imaginary parts of $k_{xj}$ enter the spinor.

Under the transformation $k_y \to -k_y$, the first component changes from $\al(k_y+ik_{xj})$ to $\al(-k_y+ik_{xj})$. When $k_{xj}$ is real, the contributions to $\langle \si_x \rangle$ from $(k_{xj},k_y)$ and $(k_{xj},-k_y)$ cancel exactly due to the opposite sign of $k_y$. However, when $k_{xj}$ is complex, the term $ik_{xj}$ also contributes to the relative phase between spinor components, and this contribution does not change sign under $k_y \to -k_y$. As a result, the expectation values $\langle \si_x \rangle$ for $(k_{xj},k_y)$ and $(k_{xj},-k_y)$ are no longer equal in magnitude and opposite in sign, leading to an incomplete cancellation. 

Furthermore, since $\Im(k_{xj})>0$ for evanescent modes, the corresponding wavefunctions decay as $e^{-\Im(k_{xj})x}$, and hence their contribution to the transverse current is localized near the interface and decreases exponentially with distance into the SOC region.

\subsection{Effect of barrier and bound states}

\begin{figure}[htb]
\includegraphics[width=4cm]{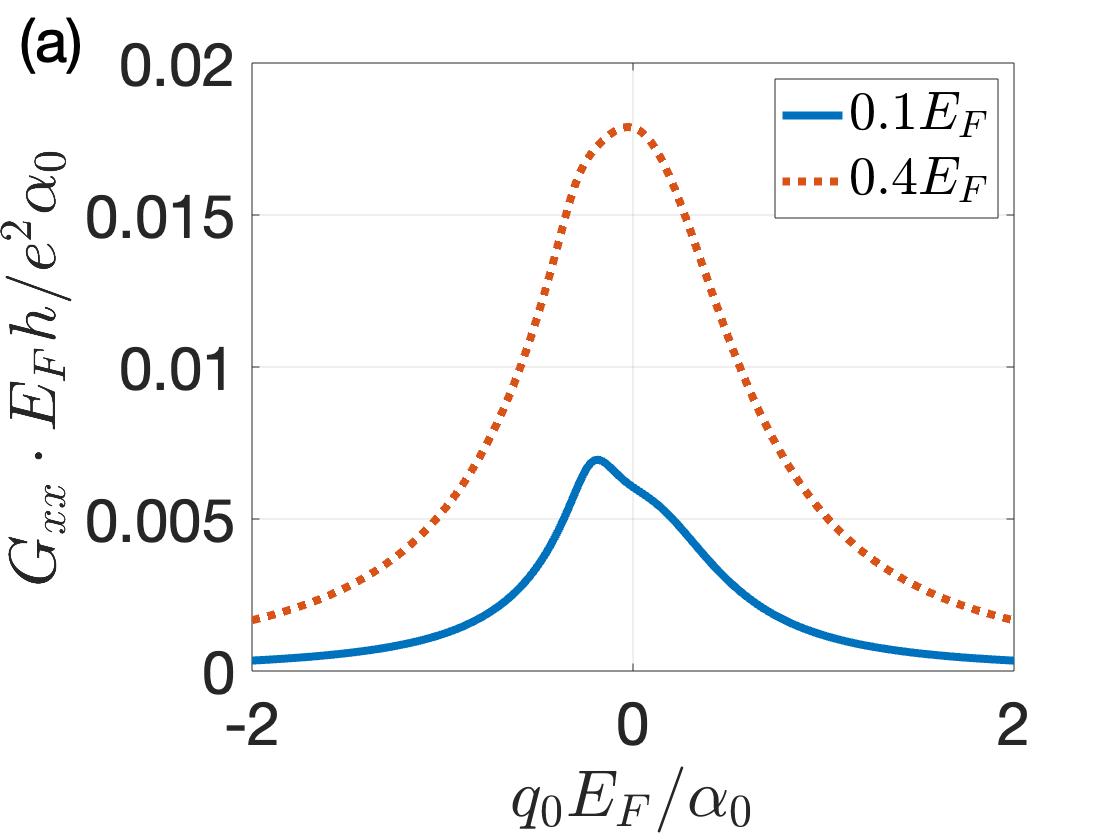}
\includegraphics[width=4cm]{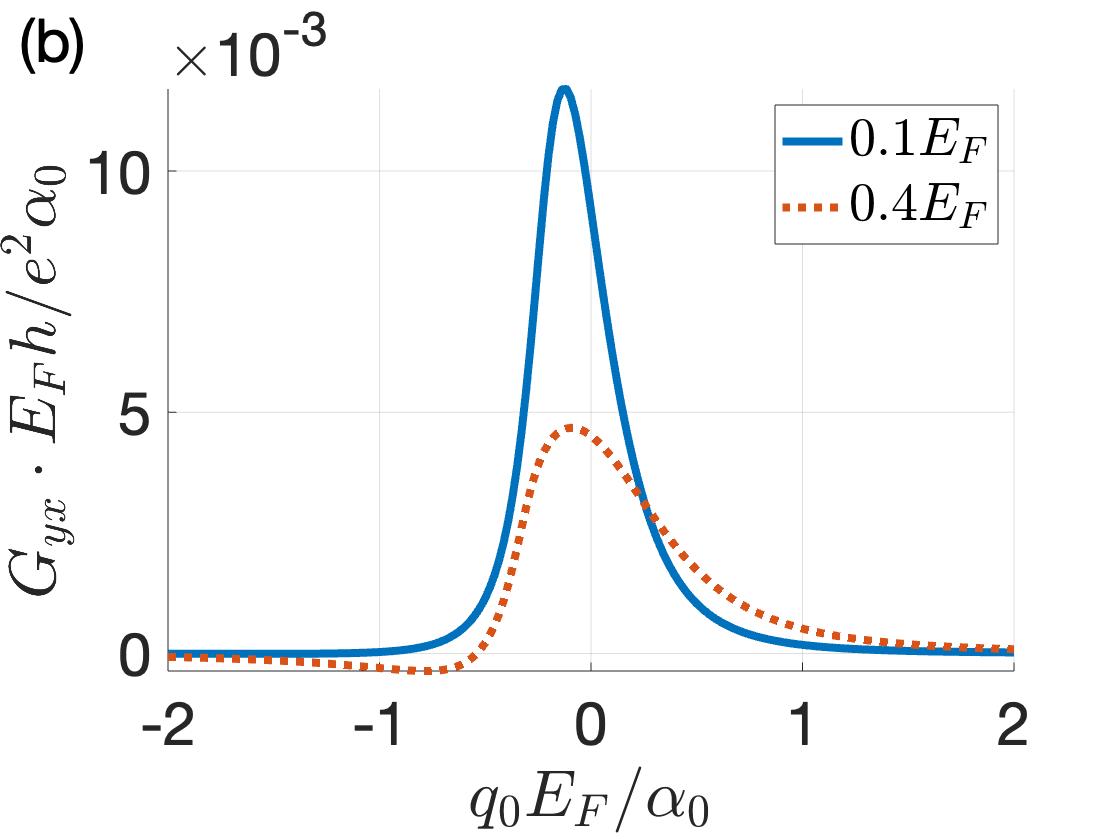}
\caption{(a) Longitudinal conductivity $G_{xx}$ and (b) transverse conductivity $G_{yx}$ versus barrier strength $q_0$ for two bias values. For small bias, the peak occurs at negative $q_0$.}
\label{fig:vsq0}
\end{figure}

Figure~\ref{fig:vsq0} shows the dependence of conductivities on the barrier strength $q_0$. For small bias ($eV=0.1E_F$), both $G_{xx}$ and $G_{yx}$ exhibit peaks at negative $q_0$, suggesting the presence of bound states at the junction. Such bound states are expected for an attractive delta-function potential. These states enhance transmission near their energy, leading to peaks in conductivity. For larger bias, the peak shifts toward $q_0=0$, as the bound-state energy lies farther from the transport window.

\subsection{Transverse response without transmission}

To further elucidate the origin of the transverse current, we consider a geometry in which the spin--orbit coupled region extends over a finite length along the $x$-direction and is terminated without a lead on the right. In this case, all incident electrons are reflected, resulting in vanishing longitudinal conductivity.

Despite the absence of transmission, we find that a finite transverse conductivity is generated within the SOC region. This demonstrates that the transverse response does not rely on net charge transport across the junction, but instead originates from the asymmetry in scattering processes at the interface. In particular,  the evanescent modes in the SOC region carry finite spin polarization and contribute to a nonzero transverse current. This provides further evidence that the mechanism underlying the transverse response is intrinsically interfacial and is closely tied to the role of evanescent modes.

\section{Bound states at the junction}

In this section, we analyze the existence of bound states localized at the junction. These states correspond to solutions of the Schr\"odinger equation that decay away from the interface on both sides and therefore do not contribute to transport, but can strongly influence it through resonant effects.

The band bottom in the NM region lies at zero energy, whereas in the SOC region it is given by $E_{sb}=-[m_s\al^2/(2\hbar^2)+\hbar^2b^2/(2m_s\al^2)]$. Bound states can therefore exist only for energies below the minimum of the continuum spectrum on both sides, i.e., for $E<E_{sb}$.

The wavefunction corresponding to a bound state must decay as $|x|\to\infty$. Accordingly, we consider the ansatz
\bea 
\psi(x) =
\begin{cases}
s_{-,\ua}e^{-ik_{x0}x}\ket{\ua}+s_{-,\da}e^{-ik_{x0}x}\ket{\da}, \\ 
\text{for}~~x<0,\\
s_{+,1}e^{ik_{x1}x}\ket{1}+s_{+,2}e^{ik_{x2}x}\ket{2},~~\text{for}~~x>0,
\end{cases}
\label{eq:psib}
\eea
where $k_{x0}=i\sqrt{k_y^2-2m_nE/\hbar^2}$ is purely imaginary so that the wavefunction decays in the NM region. Similarly, $k_{x1}$ and $k_{x2}$ are solutions of the dispersion relation in the SOC region at energy $E<E_{sb}$, chosen such that their imaginary parts are positive, ensuring decay into the SOC region.

The coefficients $s_{-,\ua}$, $s_{-,\da}$, $s_{+,1}$, and $s_{+,2}$ are determined by imposing the boundary conditions at the junction [Eq.~\eqref{eq:bc}]. These conditions lead to a homogeneous system of four linear equations, which can be written compactly as
\beq
MX = 0,
\eeq
where $X=[s_{-,\ua}, s_{-,\da}, s_{+,1}, s_{+,2}]^T$, and $M$ is a $4\times 4$ matrix whose elements depend on $E$ and $k_y$.

A nontrivial solution for $X$ exists only when the determinant of $M$ vanishes. Thus, the bound-state energies are obtained from the condition
\beq
\det M(E,k_y)=0.
\eeq

\begin{figure}[htb]
\includegraphics[width=8cm]{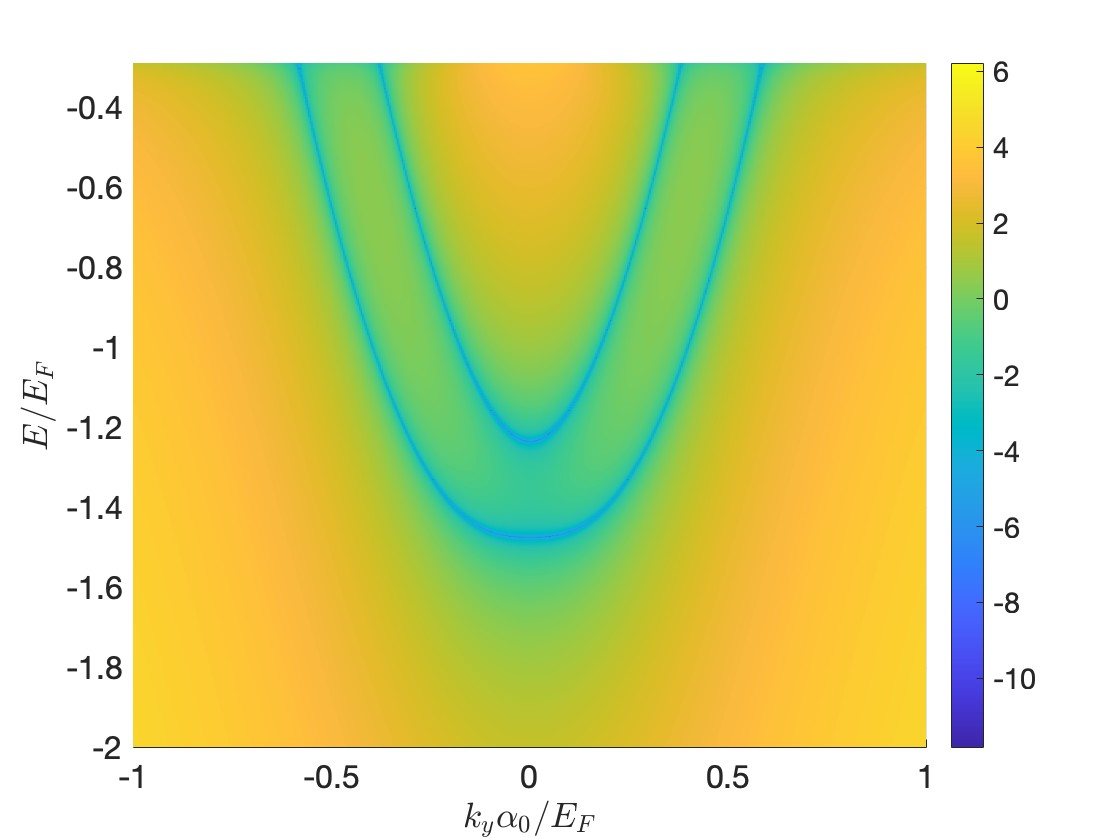}
\caption{Logarithm of $|{\rm det}M|$ plotted as a function of energy $E$ and transverse momentum $k_y$ for $q_0=-E_F/\al_0$. Bound states correspond to solutions of $\det M=0$ and appear as pronounced minima (large negative values) in the plot. The dispersion of these minima in the $(k_y,E)$ plane reflects the $k_y$ dependence of the bound-state energies. Other parameters are the same as in Fig.~\ref{fig:GvsE}.}
\label{fig:bdst}
\end{figure}

To identify these states, we compute numerically the quantity $|{\rm det}M|$ over a range of $(k_y,E)$ and plot $\log|{\rm det}M|$ in Fig.~\ref{fig:bdst} for a representative value of the barrier strength $q_0=-E_F/\al_0$. Bound states appear as sharp minima (large negative values) in this plot, corresponding to points where $\det M$ approaches zero.

The resulting dispersion of bound states in the $(k_y,E)$ plane indicates that the bound-state energy depends explicitly on the transverse momentum $k_y$. This dependence arises because $k_y$ enters the Hamiltonian as a good quantum number due to translational invariance along the $y$-direction, and the dispersion relation in both regions depends on $k_y$. Consequently, the allowed values of $k_{x0}$ and $k_{xj}$, which determine the decay of the wavefunction and enter the boundary conditions, are functions of $k_y$. This leads to a $k_y$-dependent condition $\det M(E,k_y)=0$, and hence to a $k_y$-dependent bound-state energy. These bound states play an important role in transport: when their energies lie close to the bias window, they enhance transmission through resonant processes, leading to the peaks observed in the conductivities discussed earlier.

\section{Discussion}
In Rashba metals (without an applied Zeeman field), evanescent modes are known to exhibit out-of-plane spin polarization~\cite{reynoso2006}. A related effect arises in our setup: the evanescent modes carry a finite spin polarization $\la \si_x \ra$, which in turn gives rise to a transverse current.

From an experimental standpoint, the Rashba spin-orbit coupling parameter $\alpha_R$ in InAs two-dimensional electron gases can be broadly tuned via gate voltages, with typical values ranging between $5$ and $40$~meV$\cdot$nm~\cite{matsuyama2000,Lamari2001,Lotfizadeh24,Flensberg2016}. For these experimentally realizable values of the Rashba splitting used in this paper, the ratio of the transverse conductivities in opposite bias directions,  steadily increases from zero at zero bias, crosses unity near $eV = 0.22E_F$, and reaches a value of $1.9$ at $eV = E_F$ [see Fig.~\ref{fig:GvsE}]. This indicates a robust directional asymmetry well within the resolution of standard transport measurements.

The nonzero transverse conductivity predicted in this work can be probed in a realistic setup by attaching voltage probes to the Rashba metal in the vicinity of the junction. The transverse conductivity exhibits spatial oscillations with a characteristic length scale $\sim 10E_F/\al_0 \approx 60\,{\rm nm}$. 

For bias $eV=0.4E_F$, a voltage probe with spatial resolution smaller than this length scale can resolve these oscillations and thereby measure the transverse response. In contrast, for lower bias ($eV=0.1E_F$), the transverse conductivity decays rapidly away from the junction, and hence the probe must be positioned sufficiently close to the interface to detect a measurable signal. The effect can be detected via transverse voltage measurements using nanoscale probes placed within $\sim 50$–$100\,\mathrm{nm}$ of the junction, which is well within current experimental capabilities.

In contrast to longitudinal nonreciprocal transport, which typically arises in phase-incoherent systems~\cite{khanna2024,varshney2026}, the transverse response investigated here is entirely phase-coherent.

\section{Summary and conclusions}

In this work, we have studied charge transport across a junction between a normal metal and a Rashba metal in the presence of a Zeeman field applied to the spin--orbit coupled region. While an out-of-plane Zeeman field does not generate a transverse response in a homogeneous Rashba system, we show that such a junction exhibits a finite transverse conductivity.

A key result of our work is that the transverse conductivity is \emph{nonreciprocal}, i.e., it depends on the direction of the applied bias. In particular, the transverse response for transport from left to right differs qualitatively from that for transport from right to left, with the latter exhibiting a finite transverse conductivity even at zero bias. This nonreciprocity originates from the asymmetric role of propagating and evanescent modes on the two sides of the junction.

We demonstrate that the appearance   of transverse current is rooted in the breaking of the $k_y \to -k_y$ symmetry at the level of the Hamiltonian in the presence of the Zeeman field. Consequently, the contributions to the transverse current from states with opposite transverse momenta do not cancel. We further show that evanescent modes in the spin--orbit coupled region play a crucial role: these modes carry a finite spin polarization $\la \si_x \ra$, which gives rise to a transverse current localized near the junction.

The transverse conductivity exhibits distinct signatures as a function of bias. For transport from left to right, it vanishes at zero bias, develops a peak at an energy scale set by the Zeeman field, and decreases at higher bias. In contrast, for transport from right to left, a finite transverse response appears even at zero bias due to reflected states acquiring transverse current. We also analyze the spatial dependence of the transverse conductivity and show that it decays or oscillates in the spin--orbit coupled region depending on whether the relevant modes are evanescent or propagating.

Finally, we investigate the role of a barrier at the junction and demonstrate the existence of bound states for attractive barrier strengths. These bound states disperse with transverse momentum and enhance conductivity when their energies lie near the transport window.

Our results reveal a mechanism for generating \emph{nonreciprocal transverse charge transport} in systems with Rashba spin--orbit coupling without requiring in-plane magnetic fields or ferromagnetic contacts. The predicted effects should be experimentally accessible in semiconductor heterostructures with gate-tunable Rashba coupling and Zeeman fields.

\acknowledgements 
We thank Diptiman Sen and Dhavala Suri for useful discussions.  MB and AS  thank Anusandhan National Research Foundation (erstwhile Science and Engineering Research Board) (India)  for financial support through Core Research grant (CRG/2022/004311).  BKS thanks Ministry of Social Justice and Empowerment, Government of India for fellowship through NFOBC.

\appendix 
\section{Robustness against Zeeman field extension into the normal metal}
Confining the proximity-induced Zeeman splitting strictly to the spin-orbit coupled region may present experimental challenges due to potential magnetic spillover. To address this practical concern, we assess how our results are modified when a uniform out-of-plane Zeeman field ($b$) is present across both the NM and SOC regions. This results in different dispersion relations for the electrons of two spins in NM region given by $E=\hbar^2\vec k^2/2m_n+s_{\si} b$ where $s_{\ua}=1$, $s_{\da}=-1$. 
 Consequently, for an incident electron in the spin-$\sigma$ channel from NM, the scattering wavefunction in the NM region changes from Eq.~\eqref{eq:psi} to:
 
\bea
 \psi_{\si}(x) &=& e^{ik_{x\si}}|\si\ra +r_{\ua\si} e^{ik_{x\ua}} |\ua\ra+ r_{\da\si} e^{ik_{x\da}} |\da\ra, \\
&& {\rm where} ~~k_{x\si}=\cos{\th}\sqrt{2m_n(E-s_{\si}b)}, \nn \\ && k_{y\si}=\sin{\th}\sqrt{2m_n(E-s_{\si}b)}, 
\eea
with the wavefunction for $\si$-spin incidence being $\psi_{\si}(x)e^{ik_{y\si}y}$. Here, $k_{x\bar\si}$  the wavenumber for the electrons of spin opposite to $\si$  needs to be calculated from the corresponding dispersion relation keeping $k_{y}$ the same.  Now, the expression for conductivities given by Eq.~\eqref{eq:condu} remain the same, but the expressions for the currents given by Eq.~\eqref{eq:current} change to 
\bea
I_{x} &=& \f{e}{\sqrt{m_n}}\cos{\th}\,{\rm Re}\big[\sqrt{2(E+b)}+\sqrt{2(E-b)}\nn \\ &&-\sum_{\si,\si'}\sqrt{2(E-s_{\si}b)}|r_{\si\si'}|^2\big],\\
I_{y} &=&
\begin{cases}
{e\hbar }\sum_{\si=\ua,\da}k_{y\si}\psi_{\si}^{\dagger}(x)\psi_{\si}(x)/m_n,~~~{\rm for ~~}x<0,\\~~~\\
e\sum_{\si=\ua,\da}\Big[{\hbar k_{y\si}}\psi_{\si}^{\dagger}(x)\psi_{\si}(x)/m_s \nn \\~~~~~~~~ -{\al}\psi_{\si}^{\dag}(x)\si_x\psi_{\si}(x)/\hbar\Big], 
{\rm ~~~~for~~ }x>0 .
\end{cases} 
\eea 

\begin{figure}
\includegraphics[width=8cm]{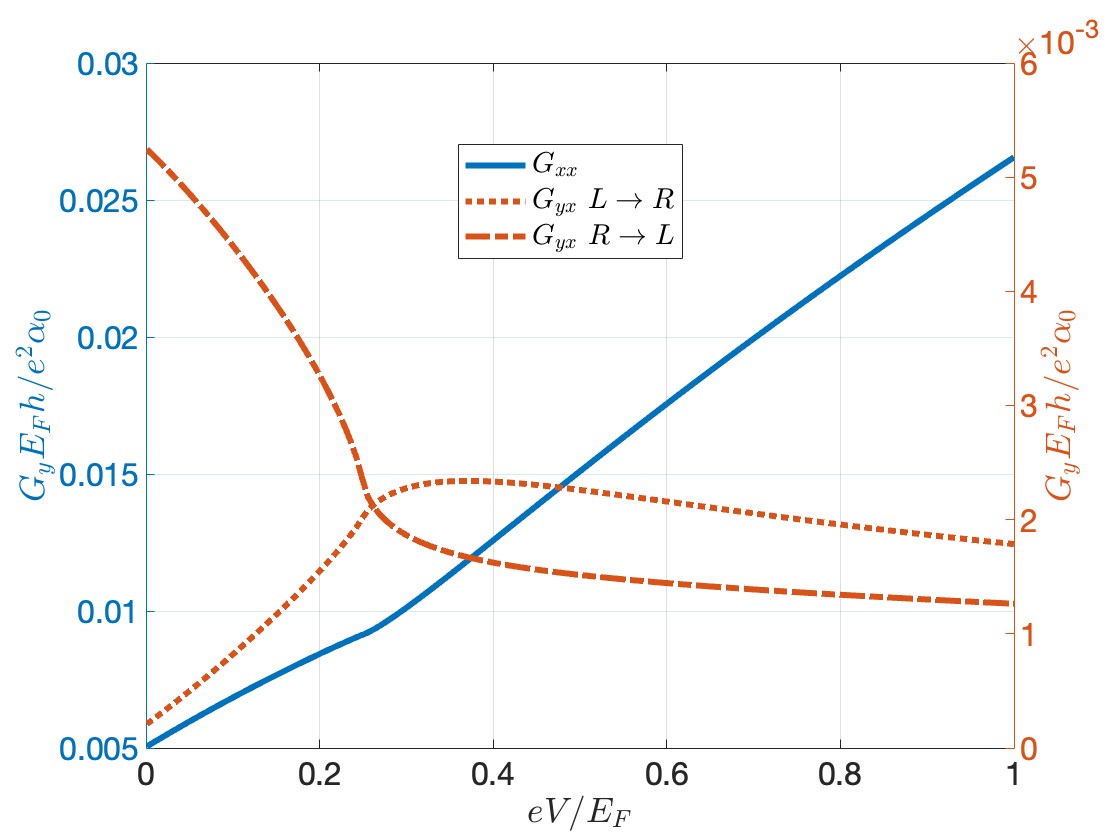}
\caption{Longitudinal ($G_{xx}$) and transverse ($G_{yx}$) conductivities as functions of bias. While $G_{xx}$ is identical for left-to-right (L$\to$R) and right-to-left (R$\to$L) bias, $G_{yx}$ differs for the two directions. Note the strong non-reciprocity in $G_{yx}$. Parameters are same as in Fig.~\ref{fig:GvsE}, except that here, the Zeeman field is present in NM and SOC -both the regions. }\label{fig:GvxE-b}
\end{figure}

The results shown in Fig.~\ref{fig:GvsE} remain qualitatively the same except for a change in magnitude of the conductivities. In Fig.~\ref{fig:GvxE-b}, we show the results for the case of uniform Zeeman field in NM and SOC regions. 

\bibliography{ref_nmsoc}

\end{document}